\documentclass[11pt,dvips]{article}
cm \voffset = -2truecm

\usepackage{graphicx}
\usepackage{amssymb}
\usepackage{latexsym}

\begin{document}
\begin{titlepage}
\setcounter{page}{1}
\renewcommand{\thefootnote}{\fnsymbol{footnote}}

\vspace{5mm}
\begin{center}

 {\Large \bf
Extended Weyl-Heisenberg algebra, phase operator, unitary
depolarizers and generalized Bell states}

\vspace{0.5cm}

{\bf M. Daoud$^1$}{\footnote {\it Facult\'e des Sciences,
D\'epartement de Physique, Agadir, Morocco; email: {\sf
daoud@pks.mpg.de}}} and {\bf E.H. El Kinani$^2$}

\vspace{0.5cm}

{\em $^1$ Max Planck Institute for Physics of Complex Systems,
N\"othnitzer Str. 38,\\
 D-01187 Dresden, Germany}\\
\vspace{0.2cm} {\em $^2$ Moulay Ismail University, Faculty of
Sciences and Technics,
 PO Box 509,\\
 Boutalamine, Errachidia, Morocco}\\

\vspace{4cm}

\begin{abstract}

Finite dimensional representations of extended Weyl-Heisenberg
algebra are studied both from mathematical and applied viewpoints.
They are used to define unitary phase operator and the corresponding
eigenstates (phase states). It is also shown that the unitary
depolarizers can be constructed in a general setting  in terms of
phase operators. Generation of generalized Bell states using the
phase operator is presented and their expressions in terms of the
elements of mutually unbiased bases are given.

\end{abstract}
\end{center}
\end{titlepage}

\newpage

\section{Introduction}

The existence of phase operator, for quantized single mode
electromagnetic field, was first postulated by Dirac \cite{Dirac}.
However, since the harmonic oscillator spectrum is infinite, the
creation and annihilation operators do not admit a naive polar
decomposition. The infinite dimensional character of the harmonic
oscillator Fock space constitutes a drawback in defining a phase
operator in a consistent way \cite{Louisell,Susskind,Carruthers}. To
overcome this problem, Pegg and Barnett considered a truncated
oscillator of finite dimension and defined the Hermitian phase
operator \cite{Pegg}. But the truncated oscillator algebra has also
one disadvantage from the algebraic point of view: its operators do
not form a closed algebra. Moreover, the truncated oscillator is not
the only way to realize the Pegg-Barnett phase operator. Indeed,
many approaches were proposed in the literature as for instance  the
refined version of phase operator  proposed in \cite{Daoud}. This
approach is based on the generalized Weyl-Heisenberg algebra which
provides finite dimensional representations and where the
hermiticity of the phase operator is automatically ensured.

In Section 1, we consider a particular, but nevertheless very large
class of extended Weyl-Heisenberg algebra. It comprises many
well-known oscillator algebras, as particular or limiting cases,
constructed in \cite{Daoud,Fu,Quesne1,Quesne2,Appl,Daoud2}. We shall
pay attention to extended Weyl-Heisenberg algebras possessing finite
dimensional representations. Based on polar decomposition of raising
and lowering generators, we construct algebraically the Hermitian
phase operator. We discuss the temporal stability of the phase
states (the eigenstates of the phase operator) and particularly the
key role of the phase parameter ensuring this temporal stability. We
study some quantum properties of the phase operator and compare with
the results obtained using the Pegg-Barnett formalism.

Section 2 of this work deals with the connection between the unitary
phase operators and the so-called unitary depolarizer operators.
This is mainly motivated by the fact that for finite dimensional
systems, unitary operations have a prominent importance in quantum
information processing. In particular, unitary depolarizer operators
are useful for quantum information processing as for instance
quantum teleportation, quantum dense coding
\cite{Hiroshima,Werner,Bowen} as well as in investigating quantum
optical systems. Note also that the notion of unitary depolarizer
operators is deeply related  to the entanglement phenomenon which
has attracted a considerable attention as a crucial resource for
quantum information processing involving higher dimensional systems.
Here, we show clearly that unitary depolarizer operators can be
constructed from the phase operator using the algebraic structures
of the extended Weyl-Heisenberg algebra. We stress that our
construction generalizes one presented recently in \cite{Ban}.

In Section 3, we show  that the generalized Bell states introduced
in \cite{Ben1} can be generated by means
 the generalized phase operator. It is important to mention that many efforts have been devoted, during the last years, to
such states to understand bipartite entanglement in higher
dimensional systems \cite{Bechmann,Bourennane,Cerf,Sych1,Sych2}. The
generalized Bell states provide a basis of maximally entangled
states in a finite dimensional Hilbert space. In this respect, we
examine interesting entanglement properties of some particular
superpositions of generalized Bell states. Finally, since mutually
unbiased bases \cite{Wootters1,Wootters2} can be derived from the
phase states \cite{Daoud} and the generalized Bell states are
defined by means of phase operator, it is natural to ask about the
connection between generalized Bell states and vectors of mutually
unbiased bases. In this sense, inspired by the results recently
presented
 in \cite{Klimov},  we derive such a relation taking
advantage of the relation, between phase states and mutually
unbiased bases, derived in \cite{Daoud}. This establishes a useful
link between mutually unbiased vectors containing single particle
information and Bell vectors reflecting the maximal bipartite
entanglement. Concluding remarks close this Letter.

\section{Extended Weyl-Heisenbeg algebra and phase states}

In this section we first introduce the extended Weyl-Heisenberg
algebra. We indicate the constraints of the structure relations to
get finite dimensional representations suitable to ensure the
unitarity of the phase operator. Then, we construct the  temporally
stable phase states. The expectation value of the Hermitian phase
operator is explicitly derived.

\subsection{Finite Fock space for extended Weyl-Heisenberg algebra}

The extended Weyl-Heisenberg algebra is an associative algebra over
the complex field $\mathbb{C}$, with generators $\{a^+ , a^- ,N \}$
and the unit ${\bf I}$ , satisfying  the structure relations
\begin{eqnarray}
[N, a^-] = -a^-,\qquad [N, a^+ ] = + a^+,\qquad
 [a^- , a^+] =  G(N),
\label{algebre}
\end{eqnarray}
where the non-negative function $G(N) = \big(G(N)\big)^{\dagger}$ is
some Hermitian analytic function of the number operator $N$. It is
clear that for $G(N) = {\bf I}$, one recovers the usual harmonic
oscillator algebra.  Also, this algebra reproduces the $W_k$ algebra
discussed in \cite{Daoud2} in the context of fractional
supersymmetric quantum mechanics of order $k$ and covers the
extended harmonic oscillator worked out in \cite{Quesne1,Quesne2}.

Let us consider the abstract Fock representation of this algebra by
means of a complete set of orthonormal states $\{ \vert n \rangle ,
n \in \mathbb{N}\}$ which are eigenstates of the number operator
$N$, $N\vert n \rangle =n \vert n \rangle$. In this representation,
the vacuum state is defined as $a^- \vert 0 \rangle =0$
 and the orthonormalized eigenstates are
constructed  by successive applications of the creation operator
$a^+$. Indeed, we define the actions of creation and annihilation
operators as
\begin{eqnarray}
 a^-\vert n \rangle =\sqrt{F(n)}  e^{{ i [F(n) - F(n-1)]\varphi}} \vert n - 1 \rangle ,\qquad
 a^+\vert n \rangle=\sqrt{F(n+1)} e^{{-i [F(n+1)- F(n)]\varphi}} \vert n+1  \rangle
\label{action sur les n}
\end{eqnarray}
where the structure function $F(.)$ is an analytic function and
should satisfy the conditions
\begin{eqnarray}
 F(0)=0 \qquad {\rm and} \qquad  F(n)>0 , \qquad  n=1,\ldots.
\end{eqnarray}
In what follows we shall denote the Fock space  as ${\cal F}$. On
this space the operators $a^+$ and $a^-$ are mutually adjoint, $a^+
= (a^-)^{\dagger}$. It is easy to check that the structure function
$F(.)$  satisfies the following recursion formula
\begin{eqnarray}
 F(n+1) - F(n) = G(n),
\end{eqnarray}
which gives by simple iteration
\begin{eqnarray}
F(n) = \sum_{m=0}^{n-1} G(m). \label{sum}
\end{eqnarray}
We restrict ourselves here to generalized oscillator algebra defined
through  structure functions obeying the condition
\begin{eqnarray}
 F(2s+1)=0,
\label{condition}
\end{eqnarray}
where $2s$ is a positive integer. We assume that $F(n)$ does not
have zeros at non-negative integers values of $n$, except $0$ and
$2s+1$. It follows that, in this case, the creation-annihilation
operators satisfy the nilpotency relations $(a^-)^{2s+1}=
(a^+)^{2s+1}=0$. This means that the corresponding representation is
$(2s+1)$-dimensional. It is interesting to note that using
(\ref{sum}), the condition (\ref{condition}) can be written as
\begin{eqnarray}
{\rm Tr}~G = 0
\end{eqnarray}
where the trace is over the $(2s+1)$-dimensional Fock space. The
structure function $F(N)$ can be factorized as
\begin{eqnarray}
F(N) =\frac{N}{2s}~(2s+1 - N)~f(N) \label{conditionF}
\end{eqnarray}
where $f(N)$ does not have vanishing eigenvalues for $ 0 \leq n \leq
2s$. Obviously, this algebra covers the usual harmonic oscillator
for $f(N) = {\bf I}$ when $s$ is large. It also includes the finite
dimensional oscillator algebra ${\cal A_{\kappa}}$ ($\kappa < 0$)
defined in \cite{Daoud} when $f(N) = {\bf I}$ and $\kappa = -1/2s$.
Of course, as evoked above, the algebra introduced here covers many
other variants of generalized harmonic oscillators, but we quote
only the ones possessing finite dimensional representations which
are interesting in the context of quantum optics.

The algebra (\ref{algebre}) defined by means of the structure
function $F(N)$ of the form (\ref{conditionF}) can be realized in
terms of the usual single photon operator. For that end, one can
write the creation and annihilation operators as
\begin{eqnarray}
a^- = b^-~\sqrt{1-\frac{N-1}{2s}}\sqrt{f(N)} \qquad a^+ =
\sqrt{f(N)}\sqrt{1 - \frac{N-1}{2s}}~ b^+
\end{eqnarray}
where $b^-$ and $b^+$ and the number operator $N = b^+b^-$ are the
usual harmonic oscillator operators. As example of the function
$f(N)$, one may choose the function
$$ f(N) ~ = ~ (b^-)^m(b^+)^m  \qquad m \in \mathbb{N}$$
introduced in \cite{Fu} to construct the multiphoton realization of
generalized oscillator algebra. For our purpose the explicit form of
the function $f(N)$ is unessential through this work (except in the
last subsection of the section 3 where we set $f(N) = {\bf I}$).

Using the algebraic structure of the generalized oscillator algebra,
one can introduce  an operator which generalizes the Hamiltonian
$a^+a^-$ for the one-dimensional harmonic oscillator. Starting from
\begin{eqnarray}
a^+a^- \vert n \rangle  = F(n) \vert n \rangle  \Longrightarrow H(N)
\equiv F(N) = a^+a^-.
\end{eqnarray}
We refer $H(N)$ to as an Hamiltonian associated with the extended
Weyl-Heisenberg algebra. The eigenvalue equation
\begin{eqnarray}
H(N) \vert n \rangle  = F(n) \vert n \rangle = \frac{n}{2s}~(2s+1 -
n)~f(n)\vert n \rangle
\end{eqnarray}
gives the energies  of a quantum dynamical system described by the
Hamiltonian operator $H(N)$. It is obvious that for $f(N) = {\bf I}
$ and $s \longrightarrow \infty$, the Hamiltonian reads
\begin{eqnarray}
H(N)   = \sum_{n=0}^{\infty} n \vert n \rangle \langle n \vert,
\end{eqnarray}
and for $f(N) = {\bf I}$ and $s$ finite, one has
\begin{eqnarray}
H(N)   = \sum_{n=0}^{2s} \frac{n}{2s}( 2s + 1 - n) \vert n \rangle
\langle n \vert,
\end{eqnarray}
which coincides with the Hamiltonian introduced in \cite{Daoud}. In
this particular case, it is interesting to stress that the ladder
operators $ a^+$ and $a^-$ can be related to Stokes operators
introduced in \cite{Luis1,Luis2} to define a unitary operator
representing the exponential of the phase difference between two
modes of the electromagnetic field. This relation may be expressed
as
$$ s_+ = \sqrt{s}~ a_+ \qquad s_- = \sqrt{s}~ a_- \qquad s_3 = \frac{1}{2}(N - s)$$
and one can simply verify that the Stokes generators $ s_+ $, $ s_-$
$s_3$ close the $su(2)$ algebra. It follows that, in the special
case where $f(N) = N(2s+1 - N)/s$, the extended Weyl-Heisenberg
generators can be realized \`a la Holstein-Primakoff by means of two
independent ordinary bosons (two electromagnetic field modes).
Remark that the parameter $s$ can be viewed as controlling the
deformation scheme. Indeed, when $s$ becomes large one has the
standard harmonic oscillator algebra. We mention that the idea of
extended Weyl-Heisenberg is mainly inspired by the polynomial
deformation of Lie algebras introduced in \cite{Higgs,Sklyanin} and
extensively discussed in the context of quantum algebras
\cite{Rocek,Bonatsos,Abdesselam}. Finally, we emphasize  that
quantum systems described by a nonlinear Hamiltonian $H(N)$ are
familiar in the context of nonlinear quantum optics as for instance
electromagnetic field propagating trough a nonlinear Kerr medium
\cite{Klimov1}.

\subsection{ Temporally stable phase states}

The Hilbert space ${\cal F}$ is $(2s+1)$-dimensional. The actions of
$a^-$ and $a^+$ on ${\cal F}$ are given by the equation (\ref{action
sur les n}) supplemented by
      \begin{eqnarray}
a^+ \vert 2s \rangle = 0,
      \end{eqnarray}
which easily follows from the calculation of $\langle 2s \vert a^-
a^+ \vert 2s \rangle$.

Let us consider the polar  decomposition of the creation $a^+$ and
annihilation $a^-$ operators:
      \begin{eqnarray}
a^- = E \sqrt{F(N)} \Leftrightarrow a^+ = \sqrt{F(N)} \left( E
\right)^{\dagger}.
      \label{decompo cas fini}
      \end{eqnarray}
It follows that the operator $E$ satisfies
      \begin{eqnarray}
E \vert n \rangle = e^{i [F(n) - F(n-1)] \varphi } \vert n-1 \rangle
      \label{action de E}
      \end{eqnarray}
for $n = 1, 2, \ldots, 2s$. For $n=0$, the action of $E$ is defined
by
            \begin{eqnarray}
E \vert 0 \rangle = e^{i [F(0)- F(2s)] \varphi} \vert 2s \rangle
            \end{eqnarray}
so that (\ref{action de E}) is valid modulo $2s+1$. It follows that
we have
      \begin{eqnarray}
\left( E \right)^{\dagger} \vert n \rangle = e^{-i [F(n+1) - F(n)]
\varphi } \vert n+1 \rangle,
      \end{eqnarray}
where $n+1$ should be understood modulo $2s+1$.  It is important to
stress that the operator $E$ is unitary. Therefore, equation
(\ref{decompo cas fini}) constitutes a polar decomposition of
creation and annihilation operators $a^+$ and $a^-$.

To construct the phase states associated with the finite dimensional
algebra under consideration, let us look to  the eigenstates of the
operator $E$. For this, one has to solve the eigenvalue equation
      \begin{eqnarray}
E \vert z \rangle = z \vert z \rangle , \qquad \vert z \rangle =
\sum_{n = 0}^{2s} C_n z^n \vert n \rangle
      \end{eqnarray}
with $z \in \mathbb{C}$. According to the method developed in
\cite{Daoud}, it is easy to see that the eigenvalues $z$ should
satisfy the  discretization condition
      \begin{eqnarray}
z^{2s+1} = 1,
      \end{eqnarray}
and subsequently the complex variable $z$ is a root of unity given
by
      \begin{eqnarray}
z = q^m \qquad m = 0, 1, \ldots, 2s,
      \end{eqnarray}
where
\begin{eqnarray}
q := e^{2 \pi i / (2s+1)}. \label{definition of q}
\end{eqnarray}
Then,  the normalized eigenstates $\vert z \rangle \equiv \vert m ,
\varphi \rangle$ of $E$ are given by
\begin{eqnarray}
\vert m , \varphi \rangle = \frac{1}{\sqrt{2s+1}} \sum_{n=0}^{2s}
e^{-i F(n) \varphi} q^{mn} \vert n \rangle.
\label{coherentstatemvarphi}
\end{eqnarray}
The states $\vert m , \varphi \rangle$, labeled by the parameters $m
\in \mathbb{Z}/(2s+1)\mathbb{Z}$ and $\varphi \in \mathbb{R}$,
satisfy
\begin{eqnarray}
E \vert m , \varphi \rangle = e^{i\theta_m} \vert m , \varphi
\rangle \qquad \theta_m = m \frac{2 \pi}{2s+1},
\label{operateurtheta}
\end{eqnarray}
which reflects that $E$ is indeed a phase operator. In the
particular case $\varphi = 0$, the states $\vert m , 0 \rangle$
correspond to an ordinary discrete Fourier transform of the basis
$\{ \vert n \rangle : n = 0, 1, \ldots, 2s \}$ of the
$(2s+1)$-dimensional space ${\cal F}$. At this level, we would like
to emphasize the key role of the parameter $\varphi$ introduced from
the beginning in the definition of actions of creation and
annihilation operators (\ref{action sur les n}). First, it ensures
the temporal stability of the phase states $\vert m , \varphi
\rangle$  under ``time evolution'':
\begin{eqnarray}
e^{-i H(N) t} \vert m , \varphi \rangle = \vert m , \varphi + t
\rangle
\end{eqnarray}
for any value of the real parameter $t$. Secondly, if one ignores
the parameter $\varphi$,i.e. $\varphi = 0$, the phase states
(\ref{coherentstatemvarphi}) reduce to ones derived by Pegg and
Barnett using the truncated harmonic oscillator \cite{Pegg}. This
means that for two different  extended Weyl-Heisenberg algebras
characterized by different structure functions, we will end up with
the same phase states. Then to differentiate between them, the phase
parameter is necessary. In the same spirit, it must be noticed that
the $SU(2)$ phase states obtained in \cite{Vourdas} are similar to
ones obtained by Pegg and Barnett despite the fact that the involved
symmetries are different. Hence, to avoid such a problem, we
introduced  the parameter $\varphi$ which
 allows us to keep the trace of the symmetry and the dynamics of the system under consideration. Note
also that the parameter $\varphi$ plays a key role in relating phase
states and mutually unbiased bases \cite{Daoud}. The phase states
have the following remarkable properties. For fixed $\varphi$, they
satisfy the equiprobability relation
\begin{eqnarray}
| \langle n \vert m , \varphi \rangle | = \frac{1}{\sqrt{2s+1}}
\qquad n, m \in \mathbb{Z}/(2s+1)\mathbb{Z}. \label{computational et
MUB}
\end{eqnarray}
They constitute an  orthonormal set, for fixed $\varphi$,
      \begin{eqnarray}
\langle m , \varphi \vert m' , \varphi \rangle = \delta_{m,m'}
\qquad m, m' \in \mathbb{Z}/(2s+1)\mathbb{Z}
      \end{eqnarray}
and satisfy the closure property
      \begin{eqnarray}
\sum_{m = 0}^{2s} \vert m , \varphi \rangle \langle m , \varphi
\vert = I.
      \end{eqnarray}
Finally, the overlapping between two phase states $\vert m' ,
\varphi' \rangle$ and $\vert m , \varphi \rangle$ is given by
\begin{eqnarray}
\langle m , \varphi \vert m' , \varphi' \rangle = \frac{1}{2s+1}
\sum_{n=0}^{2s} q^{\rho(m-m', \varphi - \varphi', n)},
   \label{overlap}
   \end{eqnarray}
where
         \begin{eqnarray}
\rho(m-m', \varphi - \varphi', n) = - (m - m')n + \frac{2s+1}{2\pi}
(\varphi - \varphi') F(n)
         \end{eqnarray}
and $q$ is defined in (\ref{definition of q}). This shows  that the
temporally stable phase states are not all orthogonal.

\subsection{ Expectation value of the phase operator}

In the previous subsection, we defined the unitary phase operator
and the phase states. We shall calculate the expectation value of
the Hermitian phase operator $\Theta$ defined by
\begin{eqnarray}
E = e^{i \Theta}.
\end{eqnarray}
It is clear that the Hermitian phase operator can be written as
\begin{eqnarray}
 \Theta = \sum_{m=0}^{2s} \theta_m \vert  m , \varphi \rangle \langle  m , \varphi \vert
\end{eqnarray}
where the summation is modulo $2s+1$. Using the expression of the
phase states (\ref{coherentstatemvarphi}), one obtains the expansion
of the Hermitian phase operator, in the Fock states basis, as
\begin{eqnarray}
 \Theta = \frac{2\pi}{(2s+1)^2}\sum_{n=0}^{2s} \sum_{n'=0}^{2s}    e^{-i [F(n) - F(n')] \varphi } ~ \theta(n,n')~ \vert  n \rangle \langle n' \vert
\label{operateurtheta}
\end{eqnarray}
where the objects $\theta(n,n')$ are given by
\begin{eqnarray}
 \theta(n,n')  =  \sum_{m=1}^{2s+1}~ m~ q^{m(n-n')} = (2s+1) ~\bigg[  s ~ \delta_{n,n'} +  \frac{q^{n-n'}}{ 1 - q^{n-n'}}~ ( 1 - \delta_{n,n'} ) \bigg].
\end{eqnarray}
Using this result, we are now in position to compute the
phase-number commutator. In the number basis, it can be expressed as
\begin{eqnarray}
 [\Theta , N ]  = \frac{2\pi}{2s+1} \sum_{n\neq n'}^{2s}  ~ \bigg[  \frac{(n-n')q^{n-n'}}{ q^{n-n'} - 1} \bigg] ~  e^{-i [F(n) - F(n')] \varphi } ~ \vert  n \rangle \langle n' \vert.
\end{eqnarray}
This equation shows that the diagonal matrix elements of the
phase-number commutator vanish and coincide with the result obtained
by Pegg and Barnett \cite{Pegg} when $\varphi = 0$. This indicates
clearly that the generalized Weyl--Heisenberg algebra gives an
alternative way to deal with finite dimensional harmonic oscillator
in investigating the phase properties of  an electromagnetic field.
 In this respect, the form of the phase states suggests us to consider a normalized pure state of the form
\begin{eqnarray}
\vert \Phi \rangle = \sum_{n=0}^{2s}  \Phi_n e^{in\alpha} e^{-i
F(n)\varphi} \vert n \rangle
\end{eqnarray}
to evaluate the phase probability distribution and the phase
expectation value. The parameter $\alpha$ is real and  the
coefficients $\Phi_n$ are real and positive satisfying
$$ \sum_{n=0}^{2s}  \Phi_n^2 = 1.$$
The vector $\vert \Phi \rangle$ is similar to the so-called partial
phase state introduced in \cite{Pegg} (up to phase factor involving
$\varphi$) and the phase probability distribution is
\begin{eqnarray}
\vert \langle m , \varphi \vert \Phi \rangle \vert^2 =
\frac{1}{2s+1} + \frac{2}{2s+1} \sum_{n>n'} \Phi_n \Phi_{n'} \cos
[(n-n')(\alpha - \theta_m)].
\end{eqnarray}
It is bounded as follows
\begin{eqnarray}
  \frac{1}{2s+1} \bigg[ 2 - \bigg( \sum_{n=0}^{2s} \Phi_n \bigg)^2 \bigg]  ~ \leq ~ \vert \langle m , \varphi \vert \Phi\rangle\vert^2 ~ \leq ~
 \frac{1}{2s+1} \bigg( \sum_{n=0}^{2s} \Phi_n \bigg)^2 .
\end{eqnarray}
Using (\ref{operateurtheta}), the expectation value of the Hermitian
phase operator $\Theta$ reads
\begin{eqnarray}
\langle \Theta \rangle = \langle \Phi \vert \Theta \vert \Phi
\rangle = \frac{2\pi}{(2s+1)^2}\sum_{n=0}^{2s} \sum_{n'=0}^{2s}
\Phi_n \Phi_{n'} e^{i\alpha(n-n')}\theta~(n, n')
\end{eqnarray}
which can be written as a sum of two terms
\begin{eqnarray}
\langle \Theta \rangle = \langle \Theta \rangle_{\rm d} + \langle
\Theta \rangle_{\rm nd}
\end{eqnarray}
where $\langle \Theta \rangle_{\rm d}$ stands for diagonal
contributions $(n = n')$ and $\langle \Theta \rangle_{\rm nd}$ is
the contribution from the off-diagonal terms. These quantities are
given by
\begin{eqnarray}
\langle \Theta \rangle_{\rm d} = \frac{2\pi s}{2s+1},
\end{eqnarray}
and
\begin{eqnarray}
\langle \Theta \rangle_{\rm nd} = \frac{2\pi}{2s+1}~ \sum_{n>n'}
\Phi_n \Phi_{n'} \frac{\sin((\alpha -
\frac{\pi}{2s+1})(n-n'))}{\sin(\frac{\pi}{2s+1}(n-n'))}.
\end{eqnarray}
It is clear that for $\alpha = \pi/(2s+1)$, the off-diagonal
contribution vanishes and the Hermitian phase operator expectation
value is $2\pi s/(2s+1)$.

\section{  Unitary depolarizer operators and generalized Bell states}\

As mentioned in the introduction the  extended Weyl--Heisenberg
algebra  not only introduces mathematical tools for the analysis of
finite quantum systems but also can be applied, for instance, to the
context in quantum information processing via the so-called the
unitary depolarizer operators and the generalized Bell states. In
this section, we shall present the derivation of unitary depolarizer
by means of phase operator. We also obtain generalized Bell states
via the action of unitary phase operator and  establish a relation
between them and vectors generating mutually unbiased bases (MUB).

\subsection{ Unitary depolarizers operators}

In a $d$-dimensional Hilbert space, the unitary depolarizers on a
domain ${\cal D}$ are the elements of set
\begin{eqnarray}
S_d = \{ U_{\alpha}~ \vert ~ U_{\alpha}U_{\alpha}^{\dagger} =
U_{\alpha}^{\dagger}U_{\alpha} = {\bf I}, ~ \alpha \in {\cal D}\},
\label{def1}
\end{eqnarray}
which satisfy the relation
\begin{eqnarray}
 \frac{1}{d} \sum_{\alpha \in {\cal D}} U_{\alpha} O  U_{\alpha}^{\dagger} ~ = ~({\rm Tr}O)~{\bf I}
\label{def2}
\end{eqnarray}
for any operator $O$ defined on the Hilbert space. The phase
operator associated with extended  Weyl-Heisenberg algebras
discussed in the previous section offers a simple way to define
unitary depolarizers. For this, we rewrite the unitary phase
operator $E$  as
\begin{eqnarray}
E = \sum_{n = 0}^{2s} e^{i(F(n+1)-F(n))\varphi}\vert n \rangle
\langle n+1\vert,
\end{eqnarray}
from which one can obtain
\begin{eqnarray}
E^k = \sum_{n = 0}^{2s} e^{i(F(n+k)-F(n))\varphi}\vert n \rangle
\langle n+k\vert  \qquad {\rm modulo}~ (2s+1).
\end{eqnarray}
The phase states (\ref{coherentstatemvarphi}) are eigenstates of the
operator $E^k$. Indeed, one has
\begin{eqnarray}
E^k \vert m , \varphi \rangle = e^{ik\theta_m} \vert m , \varphi
\rangle.
\end{eqnarray}
Any operator $O$, acting on the Fock space ${\cal F}$, can be
expanded in the phase states basis $\{ \vert m , \varphi \rangle,
m\in \mathbb{Z}/(2s+1)\mathbb{Z} \}$ as follows
\begin{eqnarray}
O = \sum_{m=0}^{2s} \sum_{m'=0}^{2s} O_{m,m'} \vert m , \varphi
\rangle  \langle m' , \varphi \vert.
\end{eqnarray}
It is simple to check that
\begin{eqnarray}
\sum_{k=0}^{2s} E^k O {E^k}^{\dagger} =  (2s+1) \sum_{m=0}^{2s}
O_{m,m} \vert m , \varphi \rangle \langle  m , \varphi \vert.
\label{sommesurE}
\end{eqnarray}
Next, we define the unitary operator
\begin{eqnarray}
V(\varphi) = e^{i (F(N) + N)\varphi}
 \label{operatorF}
\end{eqnarray}
in terms of the number operator and the structure function $F(.)$.
Using the equations (\ref{sommesurE})-(\ref{operatorF}), the
expression of the phase states (\ref{coherentstatemvarphi}) and
performing integration over $\varphi$, one gets
\begin{eqnarray}
\frac{1}{2\pi} \int_{-\pi}^{+\pi} d\varphi \sum_{k=0}^{2s}
V(\varphi) E^k O {E^k}^{\dagger}V(\varphi)^{\dagger} = {\rm Tr}O~
{\bf I}. \label{generalban}
\end{eqnarray}
This constitutes a generalization of  the definition of unitary
depolarizers operators discussed in \cite{Ban}. Then, the definition
of unitary depolarizers (eqs. (\ref{def1}) and (\ref{def2})) can be
extended to unitary operators labeled by continuous parameters
belonging to some compact domain. It is important to note that for
quantized values of the parameter $\varphi$, given by
$$ \varphi = \frac{2\pi}{2s+1} l , \qquad l \in \mathbb{Z}/(2s+1)\mathbb{Z},$$
one can verify that the discrete analogue of the equation
(\ref{generalban}) is
\begin{eqnarray}
 \sum_{k=0}^{2s}  \sum_{l=0}^{2s} V_l E^k O {E^k}^{\dagger}V_l^{\dagger} = (2s+1){\rm Tr}O~ {\bf I}
\end{eqnarray}
where $V_l \equiv V(2\pi l/(2s+1))$. This agrees with  the result
obtained in \cite{Ban}. So, the unitary depolarizer operators can be
expressed in terms of the phase operator. To close this subsection,
it is important to stress that the construction of unitary
depolarizers is done without specifying the nature of the structure
function $F(.)$. This provides us with a general scheme to define
the unitary depolarizers for a large class of extended
Weyl--Heisenberg algebras and can be adapted to quantum systems with
higher symmetries.

\subsection{Generalized Bell states and MUB}
In this last part of our work, we shall focus on extended
Weyl--Heisenberg algebra defined by the structure function
\begin{eqnarray}
F(N) = \frac{N}{2s} (2s+1 -N). \label{F}
\end{eqnarray}
In this case, we will show that the generalized Bell sates can be
generated from the maximally entangled states using the phase
operator. Recall that the generalized Bell states were first
introduced  in \cite{Ben1} to study quantum teleportation for higher
dimensional quantum systems. We discuss briefly the entanglement
properties of some particular superpositions of generalized Bell
states and  establish a connection between the generalized Bell
states and the vectors of mutually unbiased bases.

The maximally entangled state of two $(2s+1)$-dimensional systems is
\begin{eqnarray}
\vert {\rm MES} \rangle = \frac{1}{\sqrt {2s+1}} \sum_{n=0}^{2s}
\vert n \rangle\otimes\vert n \rangle. \label{mes}
\end{eqnarray}

%They are maximally entangled states as they satisfy
%\begin{eqnarray}
%{\rm Tr}_2\vert {\rm MES} \rangle \langle {\rm MES} \vert = \frac{1}{ 2s+1} {\bf I}_1.
%\end{eqnarray}
The action of the operator $E^k \otimes {\bf I}$ on the state $\vert
{\rm MES} \rangle$ leads to
%\begin{eqnarray}
%E^k \otimes {\bf I}~ \vert {\rm MES} \rangle = \frac{1}{{\sqrt 2s+1}} \sum_{n=0}^{2s} e^{-iF(n)\varphi} \vert n \rangle\otimes\vert n+k %\rangle.
%\label{bell}
%\end{eqnarray}
%For $\varphi = ????$, the states (\ref{bell}) rewrites as (???) and coincides with the generalized Bell states introduced in \cite{Planat}.
%The standard or usual expressions for Bell states can be also obtained in a similar manner. Indeed, one has
\begin{eqnarray}
E^k \otimes {\bf I}~  \frac{1}{\sqrt {2s+1}} \sum_{n=0}^{2s}  \vert
n \rangle\otimes\vert n \rangle = \frac{1}{\sqrt {2s+1}}
\sum_{n=0}^{2s}  e^{i(F(n+k)-F(n))\varphi} \vert n
\rangle\otimes\vert n + k \rangle. \label{Bell2}
\end{eqnarray}
Using (\ref{F}) and  setting $\varphi =
-\frac{2s}{2s+1}\frac{p}{k}\pi$ (this discretization condition of
$\varphi$ is similar to one introduced in \cite{Daoud} to get
mutually unbiased states from phase states), one can write the
equation (\ref{Bell2}) as
\begin{eqnarray}
E^k \otimes {\bf I}~  \frac{1}{\sqrt {2s+1}} \sum_{n=0}^{2s}  \vert
n \rangle\otimes\vert n \rangle = q^{\frac{p}{2}(k-2s-1)} \vert
\Psi_{k,p} \rangle, \label{Bell3}
\end{eqnarray}
where the states $\vert \Psi_{k,p} \rangle$ are the generalized
Bell states introduced in \cite{Ben1}. They are given by
\begin{eqnarray}
\vert \Psi_{k,p} \rangle =   \frac{1}{\sqrt {2s+1}} \sum_{n=0}^{2s}
q^{np} \vert n \rangle\otimes\vert n+k \rangle .
\end{eqnarray}
It follows  that the generalized Bell states can be obtained from
the maximally entangled states (\ref{mes}) by
means of a successive application of the unitary phase operator. \\

Finally, it is interesting to note some characteristics  of some
particular superpositions of generalized Bell states $\vert
\Psi_{k,p} \rangle$ and their expressions in terms of vectors
belonging to mutually unbiased bases.

First, it is remarkable that the superposition of the $(2s+1)^2$
maximally entangled Bell states
\begin{eqnarray}
\frac{1}{2s+1}\sum_{k=0}^{2s}\sum_{p=0}^{2s} \vert \Psi_{k,p}
\rangle =   \frac{1}{\sqrt {2s+1}} \vert 0 \rangle\otimes
\sum_{n=0}^{2s}  \vert n \rangle. \label{superposition1}
\end{eqnarray}
is a completely separable state. In other words, the superposed
state is disentangled despite that it is a superposition of
maximally entangled states. This is not always true. Indeed, the
following superposition
\begin{eqnarray}
\sum_{k=0}^{2s} \vert \Psi_{k,k} \rangle =   \frac{1}{\sqrt {2s+1}}
\sum_{n=0}^{2s}\vert n \rangle\otimes \sum_{k=0}^{2s}  q^{nk}\vert
n+k \rangle. \label{superposition2}
\end{eqnarray}
gives a maximally entangled state. The question concerning the
entanglement of superposed Bell states is very rich and deserves
more investigation. We hope to report on this issue in a forthcoming
work. There is also a nice  connection between the superposition of
Bell states and the vectors of mutually unbiased bases $B_p$
introduced in \cite{Daoud} as
\begin{eqnarray}
 B_p = \{\vert \phi_{m}^{p} \rangle =   \frac{1}{\sqrt {2s+1}} \sum_{n=0}^{2s}
 q^{\frac{p}{2}n(2s+1-n)} q^{nm} \vert n \rangle , ~ m \in \mathbb{Z}/(2s+1)\mathbb{Z}\} \qquad p \in \mathbb{Z}/(2s+1)\mathbb{Z}.
\end{eqnarray}
Indeed, the superposed states (\ref{superposition1}) can be
expressed in terms of the mutually unbiased vectors $\vert
\phi_{m}^{p} \rangle$. This expression is
\begin{eqnarray}
\frac{1}{2s+1}\sum_{k=0}^{2s}\sum_{p=0}^{2s} \vert \Psi_{k,p}
\rangle =  \vert 0 \rangle\otimes  \vert \phi_{0}^{0} \rangle,
\end{eqnarray}
and the state (\ref{superposition2}) can be written as
\begin{eqnarray}
\sum_{k=0}^{2s} \vert \Psi_{k,k} \rangle =    \sum_{n=0}^{2s}
q^{-n^2}\vert n \rangle\otimes \vert \phi_{n}^{0} \rangle.\
\label{states63}
\end{eqnarray}
The vectors $\vert \phi_{n}^{0} \rangle$ involved in the above
superposition belong to the basis $B_0$. One can go further to find
superpositions of maximally entangled states related to vectors
$\vert \phi_{n}^{p} \rangle$ with $p \neq 0$. Indeed, it is easy to
see that the action of the unitary operator ${\bf I} \otimes
e^{iF(N)\delta}$, with $\delta = \frac{2s}{2s+1}p\pi $, on the
states (\ref{states63}) gives
\begin{eqnarray}
{\bf I} \otimes q^{\frac{p}{2}N(2s+1-N)}\bigg[\sum_{k=0}^{2s} \vert
\Psi_{k,k} \rangle \bigg] =
 \sum_{n=0}^{2s} q^{-n^2}\vert n \rangle\otimes
\vert \phi_{n}^{p} \rangle.
\end{eqnarray}
This  establishes a simple relation between the generalized Bell
states and the vectors of the mutually unbiased bases. To close this
section, we notice that the vectors of mutually unbiased bases $B_p$
can be defined as the quadratic Fourier transform of the Hilbert
space basis \cite{Kibler}. Remark also that the generators of
generalized Pauli group, which constitutes an important ingredient
to deal with the relation between generalized Bell states and MUBs
\cite{Klimov}, can be defined as in \cite{Kibler}.

%\begin{eqnarray}
%q^{N^2} \otimes q^{\frac{p}{2}N(2s+1-N)}\bigg[\sum_{k=0}^{2s} \vert \Psi_{k,k} \rangle \bigg] =
% \sum_{n=0}^{2s} \vert n \rangle\otimes
%\vert \phi_{n}^{p} \rangle.
%\end{eqnarray}

\section{\bf Concluding remarks}

This Letter is based on the extended Weyl-Heisenberg algebra. This
algebra is interesting in many respects. It describes in a unified
way a large class of generalized harmonic oscillators encountered in
the literature. The extended Weyl-Heisenberg algebra, under some
assumptions, possess finite dimensional representations. The
nonlinear structure function characterizing the algebra can take
into account some nonlinear effects that may occur in the quantum
description of quantized modes of the electromagnetic field
\cite{Walls}. In connection with the extended Weyl-Heisenberg
algebra, this work deals with the construction of a unitary phase
operator and  the determination of its temporally stable eigenstates
(the so-called phase states). This construction is valid for a large
class of generalized harmonic oscillators. We also discussed the
derivation of unitary depolarizer operators from the unitary phase
operators. We have shown that the formalism, presented in
\cite{Ban}, using the Pegg-Barnett operator can be generalized to a
large variety of extended harmonic oscillator.  Finally, we
discussed the generation of generalized Bell states \cite{Ben1}
using the phase operator. We also investigated the degree of
entanglement of superpositions of Bell states and established
a connection between them and the vectors of mutually unbiased bases.\\

{\bf Acknowledgments}:

M.D.  would like to thank the hospitality and kindness extended to
him by the Max Planck Institute for Physics of Complex Systems
(Dresden, Germany) where this work was done.

%\newpage

\end{document}